\documentstyle[preprint,eqsecnum,aps]{revtex}

\begin{document}

\setlength{\baselineskip}{.1in}

\input epsf
%
\tighten

\title{%
GENERALIZED QUANTUM THEORY AND 
BLACK HOLE EVAPORATION\thanks{\setlength{\baselineskip}{.1in}
Talk presented at the {\sl $49^{\rm th}$ Yamada
Conference: Black Holes and High Energy Astrophysics},  Kyoto, Japan, April
7-10, 1998.}}

\author{James B. Hartle\thanks{hartle@cosmic.physics.ucsb.edu}}

\address{\setlength{\baselineskip}{.2in}
\sl Department of Physics,\\
University of California,\\
Santa Barbara, CA 93106-9530}

\maketitle

\date{\today}

\section*{Abstract}

Usually quantum theory is formulated in terms of the evolution of states
through spacelike surfaces.  However, a generalization of this formulation
is needed for field theory in spacetimes not foliable by spacelike
surfaces, or in quantum gravity where geometry is not definite but
a quantum variable. In particular, a 
generalization of usual quantum theory is needed for field
theory in the spacetimes that model the process of black hole evaporation.
This paper discusses a spacetime generalization of usual quantum theory
that is
applicable to evaporating black hole spacetimes.  In this generalization,
information is not lost in the process of evaporation.  Rather, complete
information is distributed about four-dimensional spacetime. Black hole
evaporation is thus not in conflict with the principles of quantum theory
when suitably generally stated.


\setcounter{footnote}{0}
\section{Prelude}
\label{sec: 1 }

It is both an honor and a pleasure to participate in this celebration of
Professor Humitaka Sato's 60$^{\rm th}$ birthday.  He is notable for his
very broad range of scientific activities and the many areas of research 
he has pioneered. In this pioneering spirit, I offer the following essay on
the connections between gravity, quantum mechanics, and black hole
evaporation.

\section{Introduction}
\label{sec: 2}

The lesson of general relativity is that spacetime geometry is four
dimensional and changing dynamically. Yet the familiar formulations of
quantum mechanics
are at variance with this lesson. They require a fixed
spacetime geometry and a division of that geometry into space and time.
This is because familiar quantum mechanics deals with states
$|\Psi(\sigma)\rangle$ defined on spacelike surfaces, $\sigma$.
The state supplies complete information about the system at any one time.
It evolves unitarily between spacelike surfaces
\newpage
\centerline{\epsfysize=2.00in \epsfbox{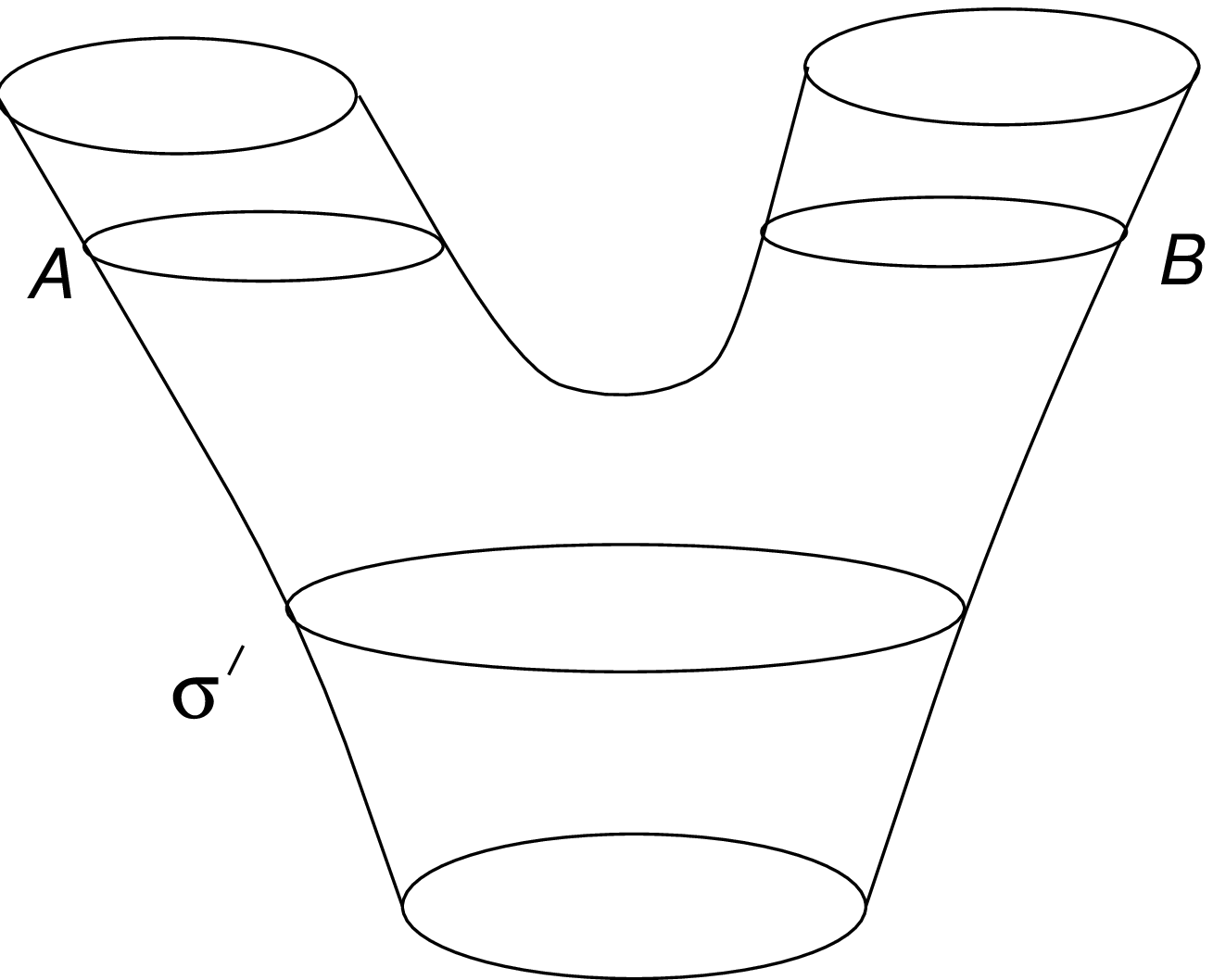}}
\vskip .13 in
\begin{quote}
{{\bf Figure 1:}
{\sl A spacetime with a simple change in spatial
topology.  There is no non-singular family of spacelike surfaces
between an ``initial''  spacelike surface $\sigma^\prime$ and ``final''
spacelike surfaces $A$ and $B$. The usual notion of quantum evolution
must therefore be generalized to apply to spacetimes such as this.}}
\end{quote}

\begin{equation}
|\Psi(\sigma^{\prime\prime})\rangle = U | \Psi (\sigma^\prime)\rangle
\label{twoone}
\end{equation}
or by reduction on them
\begin{equation}
|\Psi(\sigma)\rangle \to
\frac{P|\Psi(\sigma)\rangle}{\left\Vert P|\Psi(\sigma)\rangle\right\Vert}
\ .
\label{twotwo}
\end{equation}
Geometry must be fixed to define these surfaces and the spacetime must be
such that it can be foliated by them.  Put more technically, usual quantum
mechanics requires a fixed, globally hyperbolic spacetime. 

Even if the geometry is fixed, there are many spacetimes of interest
for which this formulation of quantum mechanics is inadequate. Take for
example, the much discussed idea of topology change illustrated in Figure 1.
There is no non-singular family of spacelike surfaces interpolating between
an initial surface $\sigma^\prime$ and final surfaces $A$ and $B$. A
generalization of familiar quantum mechanics is required for field theory
in such background spacetimes.  
Similarly, a generalization is required 
 for the spacetimes with closed timelike curves schematically
illustrated in Figure 2 which are not foliable by any family of spacelike
surfaces.

However, nowhere does the inadequacy of the usual formulations of quantum
mechanics emerge more clearly than in the process of black hole
evaporation. Black holes and quantum theory have been inextricably linked
since Hawking's 1974 discovery of the tunneling
radiation from black holes that bears
his name \cite{Haw74,Haw75}. Spacetime geometries
representing a process in which a black hole forms and evaporates
completely have global structures summarized by the Penrose diagram in
Figure 3.  Let us consider for a moment the problem of quantum mechanics
of matter fields in a fixed geometry with this global structure. 
Any pure initial
state
$|\Psi(\sigma^\prime)\rangle$
leads to a state of disordered Hawking radiation on a hypersurface
$A$ after the black hole has evaporated,
so we  have
\vskip .13in
\centerline{\epsfysize=2.00in \epsfbox{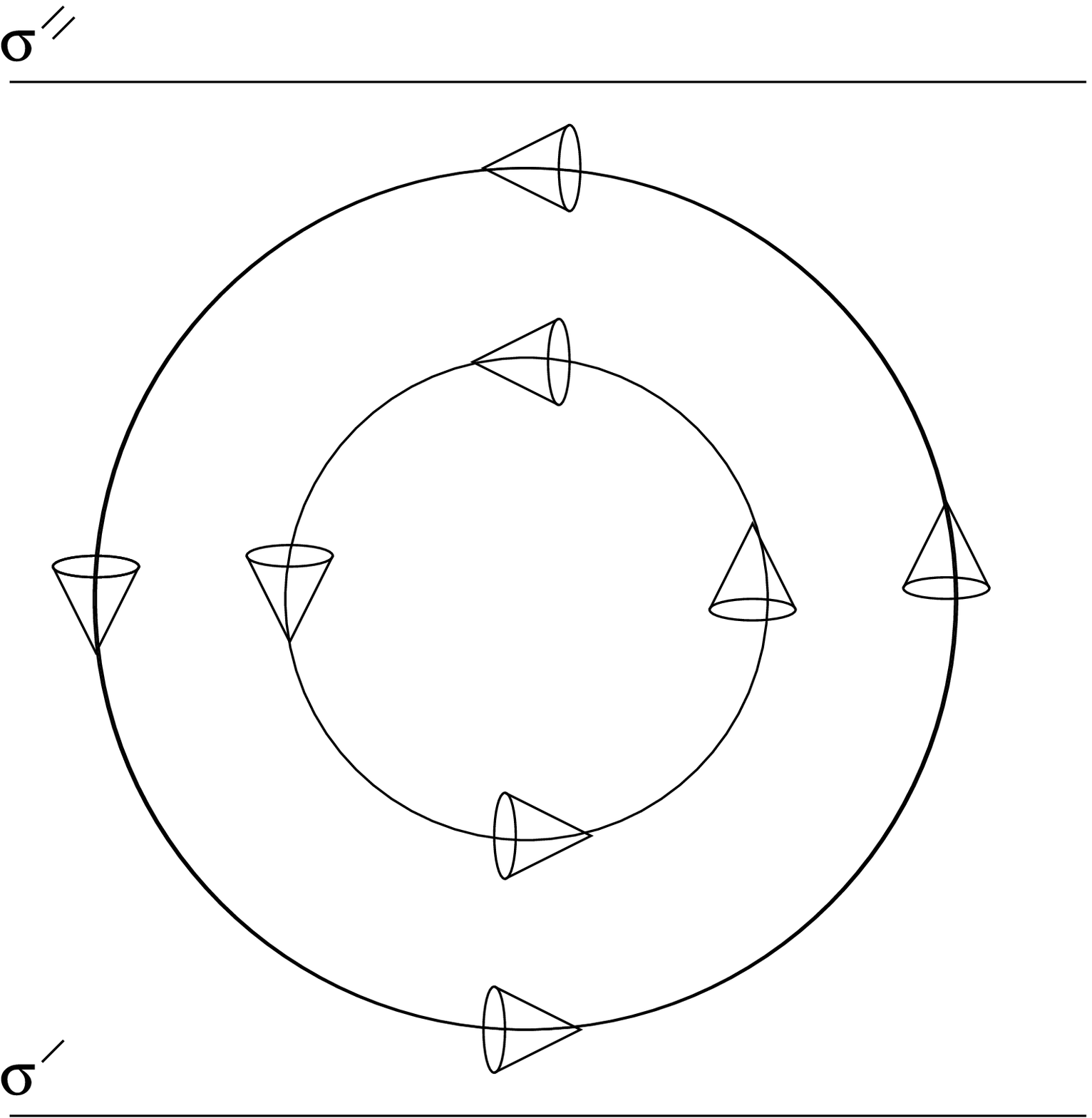}}
\vskip .13 in
\begin{quote}
{{\bf Figure 2:} {\sl
A spacetime with a compact region of closed timelike curves (CTC's).
As a consequence of the CTC's there is no family of
spacelike surfaces connecting an initial spacelike surface
$\sigma^\prime$ with a final one $\sigma^{\prime\prime}$, both outside
the CTC region.
The quantum evolution of a matter field therefore cannot be described by a
state
evolving through a foliating family of spacelike surfaces. The notion of
quantum evolution must be generalized to apply to spacetimes such as
this.}}
\end{quote}
\vskip .13 in

\begin{equation}
|\Psi(\sigma^\prime)\rangle \langle\Psi(\sigma')| \to \rho
\left(A \right)\ ,
\label{twothree}
\end{equation}
where $\rho(A)$ is the mixed density matrix describing the disordered
radiation. This transition
cannot be achieved by unitary evolution.  Indeed, it is
difficult to conceive of {\it any} law for the evolution of a
density matrix $\rho(\sigma)$ through a family of spacelike surfaces
that would
result in (\ref{twothree}) because there is no non-singular family of
spacelike surfaces that interpolate between $\sigma^\prime$ and
$A$. Even classically there is no well defined notion of evolution of
initial
data on $\sigma'$ to the surface $A$ because of the naked singularity $N$.
A generalization of quantum theory is needed for field theory in evaporating
black hole spacetimes.

The evaporating black hole spacetimes illustrated in Figure 3 are
singular. Spacetimes that are initially free from closed timelike curves
but evolve them later must be singular or
violate a positive energy condition
\cite{Tip77}. Spatial topology change implies either a singularity or
closed timelike curves \cite{Ger67}. These
pathologies suggest a breakdown in a purely classical description of
spacetime geometry.
One might therefore hope that the difficulties with usual
formulations of quantum
theory in such backgrounds could be resolved in a quantum theory of
gravity. But in any quantum
theory of gravity spacetime geometry is not fixed. Geometry is a quantum
variable --- generally fluctuating and without definite value. Quantum
dynamics cannot be defined by a state evolving in a given spacetime; no
one spacetime is given.  A generalization of usual quantum mechanics is thus
needed. This need for generalization becomes even clearer if one accepts
the hint from string theory that spacetime geometry is not fundamental.
If spacetime is not fundamental, the fundamental law of quantum evolution
cannot be the unitary progress of a state vector through surfaces {\it in}
spacetime.
\vskip .13in
\centerline{\epsfysize=3.00in \epsfbox{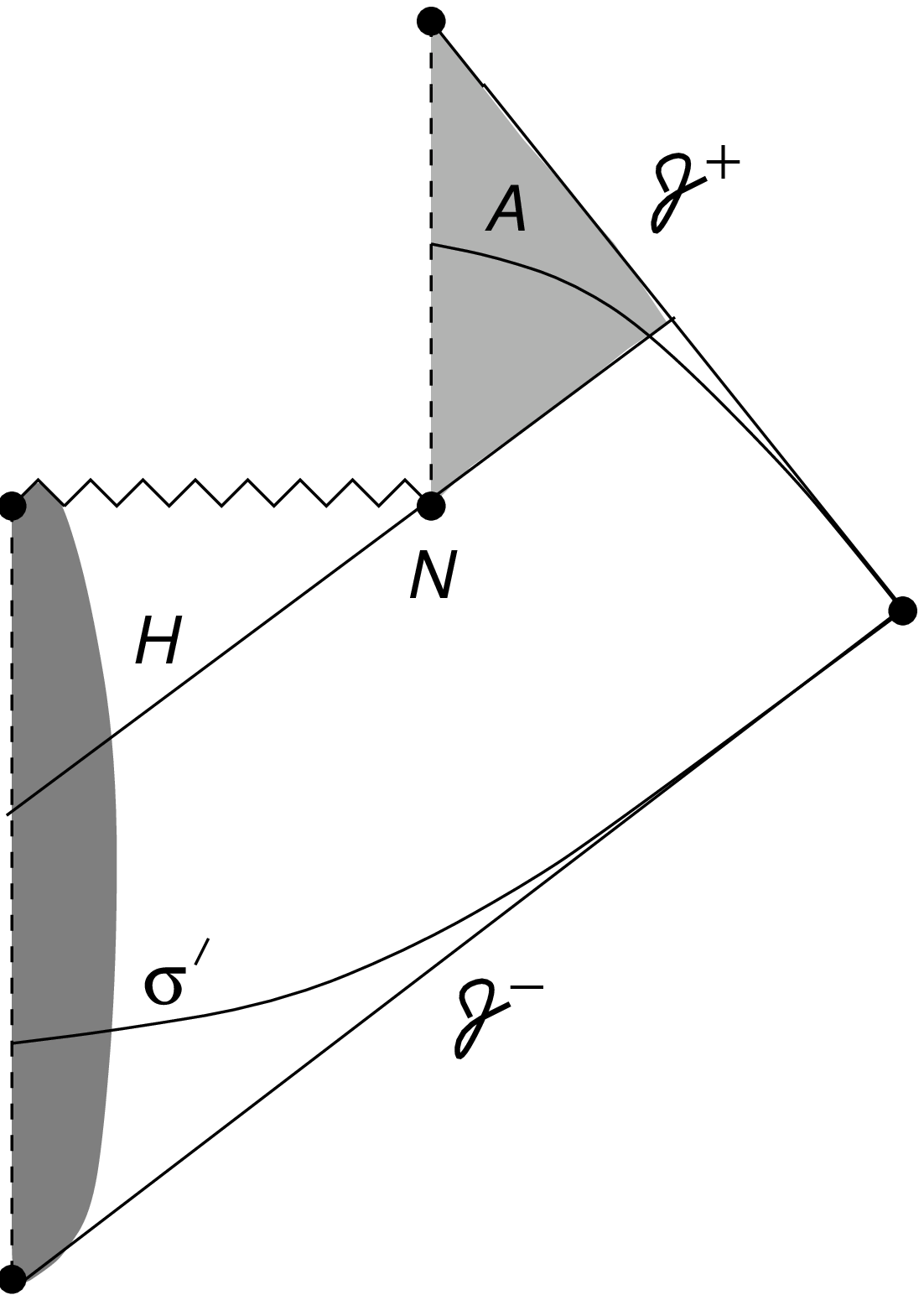}}
\vskip .13 in
\begin{quote}
{{\bf Figure 3:} {\sl The Penrose diagram for an evaporating black hole
spacetime.  The black hole is assumed here to evaporate completely, giving
rise to a naked singularity $N$, and
leaving behind a nearly flat spacetime region (lightly shaded above).
Data on  surface $A$  completely
determine the evolution of fields to its future.
Yet complete information about a quantum matter field moving in
this spacetime is not available on $A$. Even if the initial state of the
matter field on a surface like $\sigma^\prime$ is pure, the state of the
disordered Hawking radiation on $A$ would be represented by a density
matrix.  A
pure state cannot evolve unitarily into a density matrix, so the usual
formulation of quantum evolution in terms of states evolving through a
foliating family of spacelike surfaces breaks down.
The geometry of evaporating
black hole spacetimes suggests why.  There is no smooth family of
spacelike surfaces interpolating between $\sigma^\prime$ and $A$ and
even classically there is not a well defined notion of evolution
of initial data on $\sigma'$ to $A$. The usual notion
of quantum evolution must therefore be generalized to apply in spacetime
geometries such as this.}}
\end{quote}
\vskip .13 in

The conflict between the principles of general relativity and the familiar
formulation of quantum theory is usually called ``the problem of
time'' in quantum 
gravity.\footnote{For classic reviews, see Kucha\v r \cite{Kuc92}, 
Isham \cite{Ish92}, and
Unruh \cite{Unr91}.}. It
is not a problem with giving meaning to time. Rather, it is a problem with
the conflict between 
the preferred role of time (or spacelike surfaces) in usual quantum theory
and the lesson of general relativity that there are no preferred times. A
natural route forward is to generalize quantum theory so that it does not
require a preferred time, or notion of spacelike surfaces, but such that
the usual formulation is recovered approximately in the particular case when
spacetime geometry is approximately fixed and foliable by spacelike
surfaces.  

A number of such generalizations have been proposed.
Perhaps the oldest is the idea of
Dirac, Wheeler, and others that when spacetime is dynamical, quantum
mechanics should deal with wave functions defined, not on spacelike
surfaces in a fixed spacetime (there is none), but rather
on spacelike surfaces in
the superspace of three-geometrics.
This, or spaces related to it, is the space of canonical quantum gravity.
Historically, this has been an important direction for quantum gravity.
However, it is still a significant departure from the lesson of general
relativity that the fundamental notion is space{\it time}.

As stressed by Wald \cite{Wal94}
the algebraic approach to quantum mechanics 
provides a way of dealing with field theory in non-globally hyperbolic
spacetimes. Quantum dynamics is not encoded in a rule for the evolution of a
state through spacelike surfaces, but rather through a spacetime algebra of
local field operators. 
States are then defined as maps from the algebra to real numbers
({\it e.g.}, expected values). States referring to spacelike surfaces such
as $A$ in Figure 3 will naturally be mixed.

However, I do not want to discuss these particular proposals, but instead
present 
a framework for treating a {\it whole
class} of generalizations of quantum mechanics and use it to illustrate
another
way of making quantum theory consistent with the lessons of general
relativity. The idea is to follow Feynman and put quantum mechanics in fully
spacetime form; to deal with probabilities for histories rather than with
alternatives confined to spacelike surfaces; to deal with evolution 
through amplitudes for four-dimensional
histories rather than by evolving
states through spacelike surfaces. Such a formulation of quantum mechanics
is at least general enough to use for field theory in non-globally
hyperbolic spacetimes including those that describe black-hole evaporation.

To make this generalization, we will have to give up on the notion of states
on a spacelike surface and quantum evolution described by the change of
such state from one spacelike surface to another. This basic point is
already evident classically.  There is a fully
four-dimensional description of any spacetime geometry in terms of
four-dimensional manifold, metric, and field configurations.  
For globally hyperbolic geometries that
four-dimensional information can be compressed into initial data on a
spacelike surface,  which is the classical notion of state.
By writing the Einstein equation in $3+1$ form, the four dimensional
description  can be recovered by evolving the state through a
family of spacelike surfaces. However such compression is not possible
in spacetimes like the evaporating black hole spacetime illustrated in
Figure 3. Only a four-dimensional description is possible.

Analogously, there is a fully four-dimensional formulation of quantum field
theory
in background spacetime geometries in terms of Feynman's
sum-over-field-histories.  Transition amplitudes between spacelike
surfaces are specified directly from the four-dimensional action $S$ by
sums over field histories of the form
\begin{equation}
\sum\limits_{\rm histories} \exp[iS({\rm field\ history})/\hbar]\ .
\label{twofour}
\end{equation}
When the background geometry is globally hyperbolic, these transition
amplitudes between spacelike surfaces can be equivalently calculated by
evolving a quantum state through an interpolating family of spacelike
surfaces.
However, if the geometry is not globally  hyperbolic,
we cannot expect such a $3+1$ formulation of
quantum dynamics any more than we can for the classical theory.
Following
Feynman, however, we expect a fully four-dimensional spacetime formulation
of quantum theory to supply the necessary generalization 
applicable to field theory in
evaporating black hole spacetimes and to the other examples we have
mentioned.  We shall describe this
generalization and its consequences in this article.
Think four-dimensionally.
When one does there is no problem of time and no
necessary conflict between
quantum mechanics and black hole evaporation.

\section{Generalized Quantum Theory.}
\label{sec: 3}

Quantum mechanical interference between histories is 
an obstacle to thinking four dimensionally. Most generally, we
would like to predict probabilities for the individual members of sets of
alternative four-dimensional histories of a closed system.
But interference prevents the consistent assignment of probabilities to
many sets of histories.  This is
clearly illustrated in the famous two-slit experiment shown in Figure 4.
There are two
possible histories for an electron which proceeds from the source to a
point $y$ at the detecting screen,
 defined by which of the two slits (A
or B) it passes through.  It is not possible to assign
probabilities to this set of two histories.  It would be inconsistent to do so
because the probability to arrive at $y$ is not the sum of the
probabilities to arrive at $y$ {\it via} the two possible histories:
\begin{equation}
p(y) \not= p_A(y) + p_B(y)\ .
\label{threeone}
\end{equation}
That is because in quantum mechanics probabilities are squares of
amplitudes and
\begin{equation}
\left|\psi_A(y) + \psi_B(y)\right|^2 \not= \left|\psi_A(y)\right|^2 +
\left|\psi_B(y)\right|^2
\label{threetwo}
\end{equation}

In a quantum theory a rule is therefore needed to specify which sets
of alternative, four-dimensional
 histories may be assigned probabilities and which may
not.  The rule in usual quantum mechanics is that probabilities can be
assigned to the histories of the outcomes of {\it measurements} and not
in general otherwise.  Interference between histories is destroyed by
the measurement process, and probabilities may be consistently assigned.
However, this rule is too special to apply in the most general
situations and certainly insufficiently general for cosmology.
Measurements and observers, for example, were not present in the early
universe when we would like to assign probabilities to histories of
density fluctuations in matter fields or to the evolution of spacetime
geometry.

The quantum mechanics of closed systems\footnote{See, {\it
e.g.}~Omn\`es \cite{Omn94}
for a review.}
 relies on a more general rule
whose essential idea is easily stated: A closed system is in some
initial quantum state $|\Psi\rangle$. Probabilities can be assigned to
just those sets of histories for which there is vanishing interference
between the individual members 
as a consequence of the state $|\Psi\rangle$ the system is in.
Such sets of histories are said to {\it decohere}. Histories of
measurements decohere as a consequence of the interaction between the
apparatus and measured subsystem. Decoherence thus contains the rule of
usual quantum mechanics as a special case.  But decoherence is more
general.  It permits assignment of probabilities to alternative orbits
of the moon or alternative histories of density fluctuations in the
early universe when the initial state is such that these alternatives
decohere whether or not the moon or the density fluctuations
are receiving the attention of observers
making measurements.
\vskip .13 in
\centerline{\epsfysize=3.00in \epsfbox{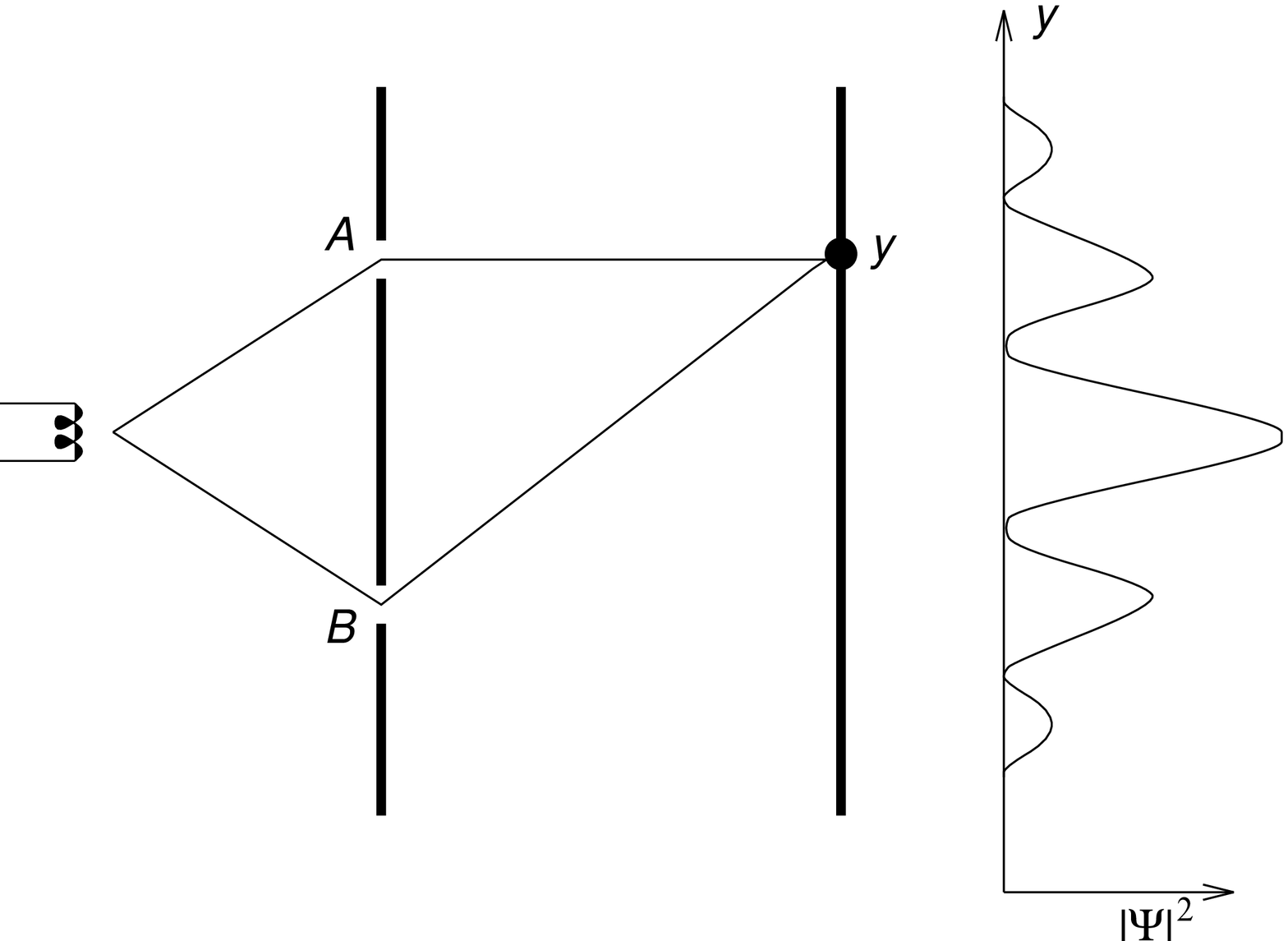}}
\vskip .13 in
\begin{quote}
{{\bf Figure 4:} {\sl The Two-Slit Experiment.
An electron gun at left emits an electron traveling towards a screen
with two slits and then on to detection at a further screen. Two
histories are possible for electrons detected at a particular point on
the right-hand screen defined by whether they went through slit A or
slit B. Probabilities cannot be consistently assigned to this set of two
alternative histories because they interfere quantum mechanically.}}
\end{quote}
\vskip .13 in

The central element in a quantum theory based on this rule is the
measure of interference between the individual coarse-grained histories
$c_\alpha, \ \alpha=1,2,\cdots$ in a set of alternative histories.  This
measure
 is
called the {\it decoherence} {\it functional},
$D(\alpha^\prime,\alpha)$. A set of coarse-grained histories decoheres when
$D(\alpha^\prime,\alpha) \approx0$ for all pairs
$(\alpha^\prime,\alpha)$ of distinct histories.
We have generally specified a generalized quantum theory when
we have specified:
\begin{itemize}
\item[(i)] The sets of alternative fine-grained histories of the system.
These are the most refined descriptions of the system
possible.  The world line of
a particle or a four-dimensional field configuration are examples.

\item[(ii)] The allowed coarse grainings of these fine-grained histories,
generally partitions of a fine-grained set into mutually exclusive classes
called coarse-grained histories. The class of world lines crossing a
particular region of space at a definite time is an example of a
coarse-grained history for a particle.

\item[(iii)]  The decoherence functional measuring the interference between
the individual members of a set of coarse-grained histories. This should
satisfy a number of general properties such as the principle of
superposition which  Isham \cite{Ish94}
has characterized axiomatically.
Usual quantum mechanics  is one
way of implementing these properties, but not the only way. Therein lie 
the possibilities for its generalization. 
\end{itemize}

\section{Spacetime Generalized Quantum Mechanics}
\label{sec: 4}

Feynman's sum-over-histories ideas may be used with the concepts of
generalized quantum theory to construct a fully four-dimensional
formulation of the quantum mechanics of a matter
field $\phi(x)$ in a fixed background spacetime. (See, {\it
e.g.}~Hartle \cite{Har95}.)
 We shall sketch this spacetime quantum
mechanics in what follows. We take the dynamics of the field to be
summarized by an action $S[\phi(x)]$ and denote the initial state of the
closed field system by $|\Psi\rangle$.

The set of fine-grained histories are the alternative, four-dimensional
configurations of the field on the spacetime. 
Sets of alternative coarse-grained histories to which the theory
assigns probabilities if decoherent are partitions of these field
configurations into exclusive classes $\{c_\alpha\}, \alpha=1,2,\cdots$.
For example, the alternative that the field configuration on a spacelike
surface $\sigma$ has the value $\chi({\bf x})$ corresponds to the class
of four-dimensional field configurations which take this value on
$\sigma$. In flat spacetime the probability of this alternative is the
probability for the initial state $|\Psi\rangle$ to
evolve to a state $|\chi ({\bf x}), \sigma\rangle$ of definite field on
$\sigma$. The history where the field takes the value $\chi^\prime({\bf
x})$ on surface $\sigma^\prime$ and $\chi^{\prime\prime}({\bf x})$ on a
later
surface $\sigma^{\prime\prime}$ corresponds to the class of
four-dimensional field configurations which take these values on
the respective surfaces, and so on.

The examples we have just given correspond to the usual quantum
mechanical notion of alternatives at a definite moment of time or a
sequence of such moments. However,
more general partitions of four-dimensional field configurations are
possible which are not at any definite moment of time or series of such
moments.  For example, the four-dimensional field configurations could
be partitioned by ranges of values of their averages over a region
extending over both space and time.  Partition of four-dimensional
histories
into exclusive classes is thus a fully four-dimensional notion of
alternative
for quantum mechanics.

Branch state vectors corresponding to individual classes $c_\alpha$
in a partition of the fine-grained
field configurations $\phi(x)$ can be constructed from the sum over
fields in the class $c_\alpha$.  We write schematically
\begin{equation}
C_\alpha|\Psi\rangle = \int_{c_\alpha}
\delta\phi\exp\Bigl(iS[\phi(x)]/\hbar\Bigr)|\Psi\rangle\ .
\label{fourone}
\end{equation}

It is fair to say that the definition of such integrals has been  little
studied in interesting background spacetimes, but we proceed assuming a
careful definition can be given even in singular spacetimes such as those
discussed earlier. Even then some discussion is needed to
explain what (\ref{fourone}) means as a formal expression. In a globally
hyperbolic spacetime we can define an operator $C_\alpha$
corresponding to the
class of histories $c_\alpha$ by specifying the matrix elements
\begin{equation}
\left\langle\chi^{\prime\prime}({\bf x}),
\sigma^{\prime\prime}\left|C_\alpha\right| \chi^\prime({\bf x}),
\sigma^\prime\right\rangle = \int_{[\chi^\prime c_\alpha
\chi^{\prime\prime}]} \delta\phi\exp\Bigl(iS[\phi(x)]/\hbar\Bigr)\ .
\label{fourtwo}
\end{equation}
\noindent The sum is over all fields in the class $c_\alpha$ that match
$\chi^\prime({\bf x})$ and $\chi^{\prime\prime}({\bf x})$ on the
surfaces $\sigma^\prime$ and $\sigma^{\prime\prime}$ respectively.  This
operator can act on $|\Psi\rangle$ by taking the inner product with its
field
representative $\langle\chi^\prime ({\bf x}),
\sigma^\prime|\Psi\rangle$ on a spacelike surface far in the past.  By
pushing $\sigma^{\prime\prime}$ forward to late times we arrive at the
definition of $C_\alpha|\Psi\rangle$. The same procedure could be used
to define branch state vectors in spacetimes with closed timelike curves
(Figure 2), in spacetimes with spatial topology change 
(Figure 1), and
in evaporating black hole spacetimes (Figure 3). The
only novelty in the latter two  cases is that $C_\alpha|\Psi\rangle$ lives
on
the product of two Hilbert spaces. There are the
Hilbert spaces on the two legs of
the trousers in the spatial topology change case and, in the black hole
case, there are the Hilbert space of states inside
the horizon and the Hilbert space of states on late time surfaces after
the black hole has evaporated.

The decoherence functional is then
\begin{equation}
D(\alpha^\prime,\alpha) = {\cal N}\langle\Psi|C^\dagger_\alpha
C_{\alpha^\prime}|\Psi\rangle\ ,
\label{fourthree}
\end{equation}
where $\cal N$ is a constant to ensure the normalization  condition
\begin{equation}
\Sigma_{\alpha \alpha'} D(\alpha', \alpha) =1  \ .
\label{new}
\end{equation}

A set of alternative histories decoheres when the
off-diagonal elements of $D(\alpha^\prime,\alpha)$ are negligible. The
probabilities of the individual histories are
\begin{equation}
p(\alpha) = D(\alpha, \alpha) = {\cal N}\left\Vert C_\alpha
|\Psi\rangle\right\Vert^2\ .
\label{fourfive}
\end{equation}
There is no issue of ``conservation of probability'' for these
$p(\alpha)$; they are not defined in terms of an evolving state vector.
As a consequence of decoherence, the $p(\alpha)$ defined by
(\ref{fourfive}) obey the most general probability sum rules including,
for instance, the elementary normalization condition $\sum_\alpha
p(\alpha)=1$
which follows from (\ref{new}).
\vskip .26 in
\centerline{
\begin{tabular}{|l|l|l|}\hline
& Usual & Generalized \\
& Quantum Mechanics & Quantum Mechanics \\\hline\hline
Dynamics & $e^{-iHt}|\Psi\rangle$
& $\sum\limits_{{\rm histories}\atop
\in c_\alpha} e^{iS({\rm
history})/\hbar}|\Psi\rangle$ \\
&$P|\Psi\rangle/\Vert P|\Psi\rangle\Vert$&\\\hline
Alternatives & On spacelike surfaces or & Arbitrary partitions of\\
& sequences of surfaces & fine-grained histories\\\hline
Probabilities & Histories of measurement & Decohering sets\\
assigned to & outcomes & of histories\\\hline
\end{tabular}
}
\vskip .26 in

This spacetime generalized quantum theory is only a modest
generalization of usual quantum mechanics in globally hyperbolic
backgrounds as the above table shows.  The two laws of evolution
(\ref{twoone}) and (\ref{twotwo})
 have been unified in a single sum-over-histories
expression. The alternatives potentially assigned probabilities have
been generalized to include ones that extend in time and are not simply
the outcomes of a measurement process.  These generalizations
put the theory in fully four-dimensional form. 

When spacetime is fixed and foliable by spacelike surfaces, this spacetime
formulation of quantum theory reduces to the familiar formulation in terms
of states on spacelike surfaces. Pick any spacelike surface $\sigma$
and define, as
in (\ref{fourone}),
\begin{equation}
\Psi \left[\chi (\bf x), \sigma\right] = \int\nolimits_{[\Psi, \chi ({\bf
x})]} \delta\phi\ \exp\left(i\, S[\phi(x)]\right)|\Psi\rangle
\label{foursix}
\end{equation}
where the integral is over field configurations to the past of $\sigma$
that match the given spatial configuration $\chi({\bf x})$ on $\sigma$. As
$\sigma$ is pushed forward in time, the resulting state vector evolves
unitarily. However, it cannot be simply pushed through a topology change, a
region of closed timelike curves, or the naked singularity in an
evaporating black hole spacetime. There the description of evolution of
states breaks down.

Thus, for coarse grainings that
are defined by alternatives on spacelike surfaces, the probabilities
predicted by this path integral formulation coincide with those predicted
by evolving states in globally hyperbolic spacetimes. But, given
appropriate definitions of the path integral, they predict probabilities
for alternatives in spacetimes with topology change, spacetimes with
closed spacelike curves, and evaporating black hole spacetimes.

Even when spacetime is not fixed but fluctuating quantum mechanically, we
can formulate a quantum mechanics of spacetime geometry within generalized
quantum theory. The histories are histories of four-dimensional geometry.
The coarse grainings are partitions of these into diffeomorphism invariant
classes, and a decoherence functional can be constructed along path
integral lines. So formulated there is no problem with time in quantum
gravity.  
Of course, $M$-theory suggests that spacetime itself is not a fundamental
notion. But even there we may expect to find notions of histories, coarse
grainings, and a theory of generalized quantum gravity free from the
problem of time.

The process of black hole evaporation is thus not in conflict with the
principles of quantum mechanics suitably generally stated.  It is not in
conflict with quantum evolution described four-dimensionally.  It is
only in conflict with the narrow idea that this evolution be reproduced
by evolution of a state vector through a family of spacelike surfaces.

\section{Information}
\label{sec: 5}

A spacetime formulation of quantum mechanics requires a spacetime notion
of information that is also in fully four-dimensional form.
In this section we describe a notion of the
information available in {\it histories} and
not just in alternatives on a single spacelike
surface.\footnote{\setlength{\baselineskip}{.1in} The
particular construction we use is due to Gell-Mann and 
Hartle \cite{G-MH90}. There are a number of other ideas, {\it
e.g.}~Isham and Linden \cite{IL97}  with which the same
points about information in a evaporating black-hole spacetime could be
made.} We then apply this to discuss the question of whether information
is lost in an evaporating black-hole spacetime.

In quantum mechanics, a statistical distribution of states is described
by a density matrix. For the forthcoming discussion it is therefore
necessary to generalize the previous considerations a bit and treat mixed
density matrices $\rho$ as initial conditions for the closed system as
well as pure states $|\Psi\rangle$. To do this it is only necessary to
replace (\ref{fourthree}) with
\begin{equation}
D(\alpha^\prime,\alpha) = Tr\left(C_{\alpha^\prime}\rho
C^\dagger_\alpha\right)\ .
\label{fiveone}
\end{equation}

A generalization of the standard Jaynes construction 
(See, {\it e.g.}~Jaynes \cite{Jay83} 
gives a natural definition of the missing information in a set of
histories $\{c_\alpha\}$. We begin by defining the
entropy functional on density matrices:

\begin{equation}
{\cal S}(\tilde\rho)  \equiv -Tr\left(\tilde\rho \log \tilde\rho\right)\ .
\label{fivetwo}
\end{equation}
With this we define the {\it missing} {\it information}
$S(\{c_\alpha\})$ in a set of histories $\{c_\alpha\}$ as the maximum
of ${\cal S} (\tilde\rho)$ over all density matrices $\tilde\rho$ that
preserve the predictions of the true density matrix $\rho$ for the
decoherence and probabilities of the set of histories $\{c_\alpha\}$. Put
differently, we
maximize ${\cal S} (\tilde\rho)$ over $\tilde\rho$ that preserve the
decoherence functional of $\rho$ defined in terms of the corresponding
class operators $\{C_\alpha\}$.
Thus,
the missing information in a set of histories $\{c_\alpha\}$ is given
explicitly by:
\begin{equation}
S(\{c_\alpha\}) =  \mathop{\rm max}_{\tilde\rho} \Bigl[{\cal S}
(\tilde\rho)\Bigr]_{\{Tr(C_{\alpha^\prime}\tilde\rho
C^\dagger_\alpha) = Tr(C_{\alpha^\prime}\rho
C^\dagger_\alpha)\}}\ .
\label{fivethree}
\end{equation}

{\it Complete} {\it information}, $S_{\rm compl}$ --- the most one can
have about the initial $\rho$ --- is found in the decoherent set of
histories with the least missing information.
\begin{equation}
S_{\rm compl} = \mathop{\rm min}_{{\rm decoherent} \atop \{c_\alpha\}}
\Bigl[S(\{c_\alpha\})\Bigr]\ .
\label{fivefour}
\end{equation}
In usual quantum mechanics it is not difficult to show that $S_{\rm
compl}$ defined in this way  is exactly the missing information in the
initial density matrix $\rho$
\begin{equation}
S_{\rm compl} = {\cal S}(\rho) = - Tr(\rho \log \rho)\ .
\label{fivefive}
\end{equation}
Information in a set of histories and complete information are {\it
spacetime} 
notions of information whose construction makes no
reference to states or alternatives on a spacelike surfaces.  Rather the
constructions are four-dimensional
making use of histories.  Thus for
example, with these notions one can capture the idea of information in
entanglements in time as well as information in entanglements in space.

One can find {\it where} information is located in
spacetime by asking what information is available from alternatives
restricted to fields in various spacetime regions.
For example, alternative values of a field average over a region $R$
refer only to fields inside $R$. The missing information in a
region $R$ is
\begin{equation}
S(R) = \mathop{\rm min}_{{\rm decohering}\{c_\alpha\}
\atop{\rm referring\ to}\ R}
S(\{c_\alpha\})\ .
\label{fivesix}
\end{equation}
When $R$ is extended over
the whole of the spacetime the missing information is complete.  But it
is an
interesting question what {\it smaller} regions $R$ contain complete
information, $S(R) = S_{\rm compl}$.

When spacetime is globally hyperbolic and quantum dynamics can be
described by states unitarily evolving through a foliating family of
spacelike surfaces, it is a consequence of the definitions
(\ref{fivefive}) and (\ref{fivesix}) that complete information is
available on each and every spacelike surface $\sigma$
\begin{equation}
S(\sigma) = S_{\rm compl} = - Tr (\rho \log \rho)\ .
\label{fiveseven}
\end{equation}
However, in more general cases, only incomplete
information may be available on any spacelike surface, and complete
information may be distributed about the spacetime.  The evaporating black
hole spacetimes in Figure 5 are an example.
\vskip .13in
\centerline{\epsfysize=3.00in \epsfbox{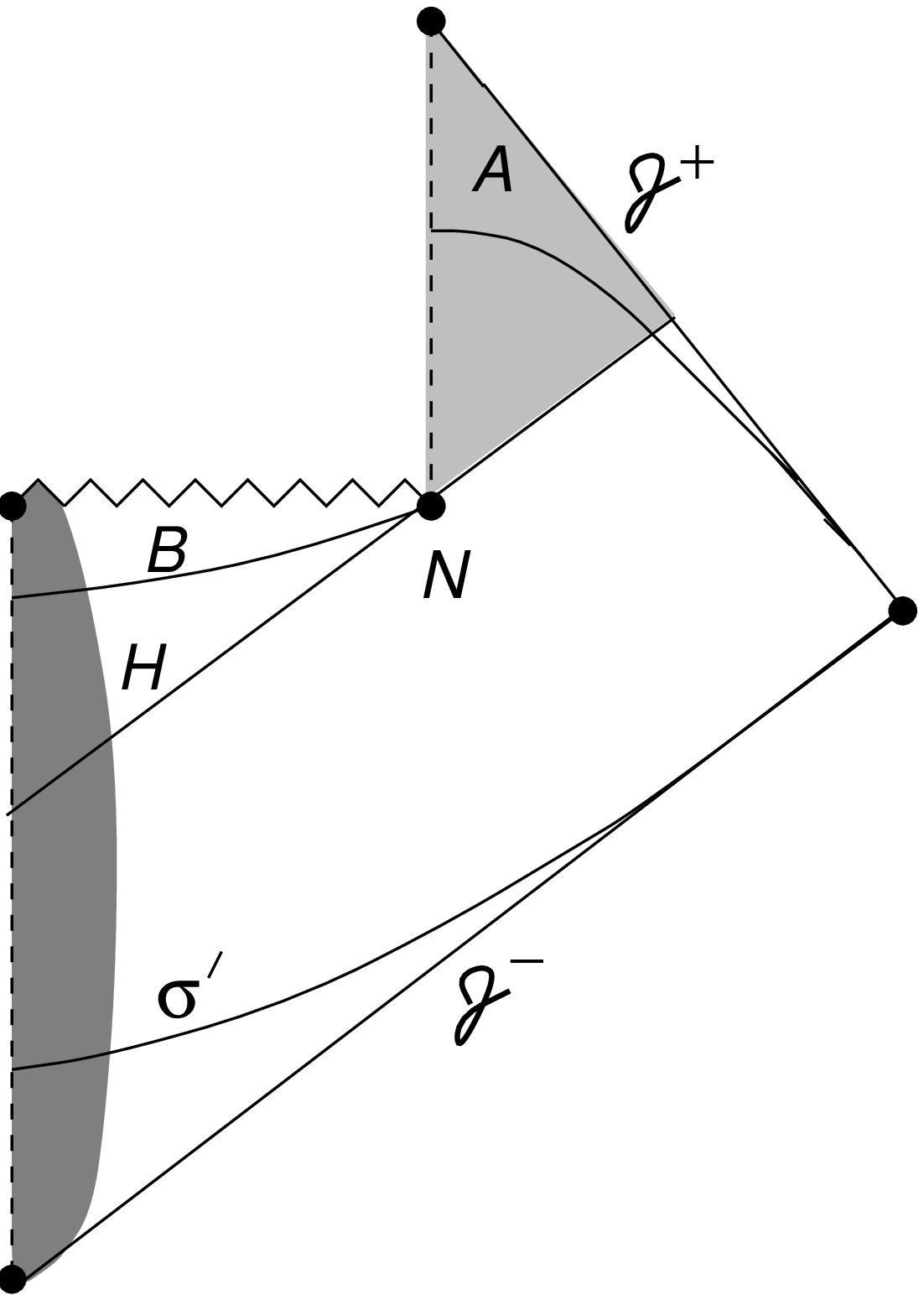}}
\vskip .13 in
\begin{quote}
{{\bf Figure 5:} {\sl Quantum evolution in an evaporating black
hole spacetime can be described four-dimensionally using a
Feynman sum-over-histories. However, that evolution is not
expressible in $3+1$ terms
by the smooth evolution of a state through spacelike surfaces.

Complete information is not recoverable on surface $A$ because of
correlations (entanglements) between the field on $A$ and the field on a
surface like $B$. Even though complete information is not necessarily
available on any one spacelike surface, it  is not lost in an
evaporating black hole geometry. It is distributed over
four-dimensional spacetime.}}
\end{quote}
\vskip .13 in

There is no reason to expect to recover complete information on
a surface like $A$ in Figure 5. One should
rather expect complete information to be available in the spacetime
region which is the union of $A$ and a surface like $B$. Even though
this region is not a spacelike surface there are still entanglements
``in time'' between alternatives on $A$ and alternatives on $B$ that
must be considered to completely account for missing information.

The situation is not so very different from that of the ``trousers''
spacetime sketched in Figure 1. There, complete information about an initial
state on $\sigma^\prime$ is plausibly not available on surfaces $A$ or
$B$ separately. That is because there will generally be correlations
(entanglements) between alternatives on $A$ and alternatives on
$B$. Complete information is thus
available in spacetime even if not available on any one spacelike
surface like $A$. The surfaces $A$ and $B$
of this example are similar in this respect
to the surfaces $A$ and $B$ of the evaporating black hole spacetime
in Figure 5.

Thus even
though it is not completely available on every spacelike surface like
$A$, information is not lost in evaporating black-hole spacetimes.
Complete information is distributed about the spacetime.

Bob Wald, who I think agrees with most of these ideas, thinks this is a bad
way of characterizing the situation.  He says that to say that information is
not lost, is like saying ``I can't find my keys which I had yesterday, but
they are not lost, they can be found in the past in spacetime''.
Information depends on coarse graining. Typically,
information {\it is} incomplete ---``lost'' --- in any given set of
coarse-grained alternatives such as those characterizing our memories or
those available to observers at infinity in an evaporating black hole
spacetime. Here, when we say information ``is not lost'', we mean that
there is some coarse-grained set of alternatives by which complete
information might have been recovered in principle.
That is the
case in black hole evaporation.

\section{Conclusion}

String theory may be a quantum theory of gravity in which there is unitary
evolution between pre-collapse and the evaporated states of a black hole.
Whether or not this is true, some
generalization of usual quantum  mechanics will be necessary when spacetime
geometry is not fixed. Further, we have sketched the principles of a
spacetime generalized quantum theory applicable both to fixed spacetime
models of black hole evaporation and more generally to quantum gravity.
Thus, black-hole evaporation is not in conflict with the principles of
quantum theory when suitably generally stated. Whether or not the
evaporation process can be described by a unitary $S$-matrix, generalized
quantum theory is ready to describe it.


\end{document}